\documentclass[article,twocolumn,eqsecnum]{revtex4-1}
\usepackage{amsmath}
\usepackage{hyperref}
\usepackage{graphicx}
\usepackage{epstopdf}
\usepackage{caption}
\usepackage{subcaption}
\usepackage{float}

\begin{document}

	\title{Quantum dynamics from fixed points and their stability}
	\author{Rohit Chawla}
	\email{rohit.chawla93@gmail.com}
	\affiliation{Department of Theoretical Physics, Indian Association for the Cultivation of Science, Jadavpur, Kolkata 700032, India}
	\author{Jayanta K. Bhattacharjee}
	\email{jayanta.bhattacharjee@gmail.com}
	\affiliation{Department of Theoretical Physics, Indian Association for the Cultivation of Science, Jadavpur, Kolkata 700032, India}	
	\date{\today}
	
	\begin{abstract}
		We approach quantum dynamics in one spatial dimension from a systematic study of moments starting from the dynamics of the mean position. This is complementary to the approach of Brizuela whose starting point was generalized recursion relations between moments. The infinite set of coupled equations is truncated which allows us to use the techniques used in the study of dynamical systems. In particular we predict for what initial variance the purely quartic oscillator will time develop with minimal change in the shape of the initial packet and what the frequency of oscillation of the mean position will be. We show how quantum fluctuations will cause a particle to escape from the well of a volcano potential and how they will cause an oscillation between the two wells of a double well potential. Further, we consider an oscillatory external field in addition to the double well potential and work near the separatrix where the classical system is known to be chaotic. We show how the quantum fluctuations suppresses the chaotic behaviour after a time interval inversely proportional to the strength of the quantum fluctuations.
	\end{abstract}
	
	\maketitle	
	
\section{Introduction}
	
	Quantum dynamics \cite{1,2,3,4} is the time evolution of an initially prescribed wave function. Given a wave function $\psi_o(x)$ at $t = 0$ for a system governed by a Hamiltonian $H$, one needs to know the wave function $\psi(x,t)$ at a later time. Since $\psi(x,t)$ can be written as $\sqrt{P(x,t)} e^{\imath \phi(x,t)}$, where $P(x,t)$ is the probability of finding the particle between $x$ and $x + dx$ at the time $t$ and $\phi(x,t)$ is the phase, we can alternatively track the probability and the phase separately. The task of finding $\psi(x,t)$ is in principle straight forward. If the relevant Hamiltonian $H$ has a complete set of eigenstates $u_n$ with energy $E_n$, then the initial state can be expanded as 
	
	\begin{equation} \label{1.1}
		\psi_o(x) = \Sigma c_n u_n(x) 
	\end{equation}	 	
	
	where the constants $c_n$ are given by $c_n = \int_{-\infty}^{\infty} \psi_o(x) u_n^*(x) dx$ and the wave function at any time $t$ is 
	
	\begin{equation} \label{1.2}
		\psi(x,t) = \Sigma c_n u_n(x) e^{-\frac{\imath E_n t}{\hbar}}
	\end{equation}	 
	
	Closed form answers can be obtained for the free particle and the simple harmonic oscillators provided $\psi_o(x)$ is a Gaussian wave packet. For other potentials(as well as non Gaussian initial wave packets for the above two) the prescription of Eq.(\ref{1.2}) is simply a numerical recipe.
	
	A different approach to dynamics is that due to Ehrenfest \cite{5,6,7,8} which somehow got restricted to an understanding of the classical limit. However, the Heisenberg equation of motion for an arbitrary operator $\hat{O}$ is purely quantum dynamics and is given by 
	
	\begin{equation} \label{1.3}
		\imath \hbar \frac{d}{dt}\hat{O} = \imath \hbar \frac{\partial}{\partial t}\hat{O} + [\hat{O},H]
	\end{equation}
	
	Taking expectation values, we have
	
	\begin{equation} \label{1.4}
		\imath \hbar \frac{d}{dt}\langle \hat{O} \rangle = \imath \hbar \frac{\partial}{\partial t}\langle \hat{O} \rangle + \langle [\hat{O},H] \rangle
	\end{equation}
	
	The above equations were frequently explored \cite{9,10,11,12,13} two to three decades ago to study the quantum dynamics of classical chaotic systems but there were hardly any definite results. Recently an interesting study of the higher order moments was conducted by Brizuela \cite{16,17}. Some generalised uncertainty relations were found and specific results obtained about the dynamics of the quartic oscillator. Systematically using Eq.(\ref{1.4}) for these moments we find an analytic picture of quantum dynamical energies for systems which hitherto were restricted to a purely numerical analysis. Our study is based on the following observations.
	
	\vspace{.1in}	
	
	\rm{i}) For a Hamiltonian having terms higher than the quadratic in the potential, the dynamics of the mean position is not the classical dynamics but is driven by quantum fluctuations expressed through variance, skewness, kurtosis etc of the probability distribution.
	
	\vspace{.1in}	
	
	\rm{ii}) Repeated use of Eq(\ref{1.4}) allows one to write down the dynamics of the various moments. Our approach to the dynamics also uses the different moments but is complementary to that of Brizuela \cite{16,17} in that we start from the lower one.
	
	\vspace{.1in}	
	
	\rm{iii}) The system of equations is not closed and requires one of the closure schemes that is common in other areas \cite{18,19,20,21,22}.
		
	\vspace{.1in}	
	
	\rm{iv}) Closure leads to a finite dimensional dynamical system which can now be subjected to the usual analysis involving fixed points and their stability \cite{23} and also can be numerically integrated much faster. 
	
	\vspace{.1in}	
	
	In Sec. II, we take up the purely anharmonic oscillator. In particular we treat the quartic oscillator with the potential $V(x) = \frac{\lambda}{4} x^4$ and explore initial states which can time develop in an approximately shape invariant manner. We predict the initial width of such states and the frequency of the motion of the centre of the wave packet. In Sec. III, we treat a volcano potential and show that our technique can allow a particle to escape from the well. The simplest scheme which uses a Gaussian approximation allows escape from the top half of the confining well but truncations at later orders allows the lower energy particles to escape as well. In Sec. IV, we treat the case of tunnelling in a double well potential and show that a Gaussian approximation allows the particles to tunnel and oscillate between the two wells if the energy is in the top one third of the well. Once again higher order truncations lead to better results.
	
	\vspace{.1in}	
	
	As a further demonstration of the efficiency of this method of studying the dynamics, we consider the double well perturbed by a small linear potential oscillating in time. The classical system is known to exhibit chaotic behaviour if the energy is such that the motion is near the separatrix \cite{24,25}. We show that the inclusion of quantum fluctuations causes the dynamics to become regular at long enough times as is to be expected \cite{26,27,28,29}.		  
	
\section{The Quantum Anharmonic Oscillator}

	We begin by considering the dynamics in the anharmonic potential(we set $m=1$),
	
	\begin{figure}[h]
		\includegraphics[width=\linewidth]{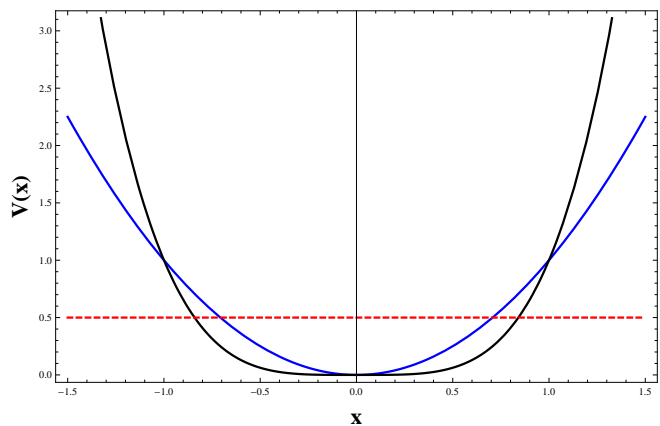}	
		\caption{(color online) Comparison of the quartic with the quadratic potential.(Black curve is the quartic term $x^4$, blue curve is the quadratic term $x^2$, red dashed line is energy $e = .5$ used in the numerical plot for $\langle x \rangle$ in Fig.(\ref{Fig. 2}))}
		\label{Fig. 1}
	\end{figure}
	
	\begin{equation} \label{2.1}
		V(x) = \frac{1}{2}\omega^2 x^2 + \frac{\lambda}{4}x^4
	\end{equation}

	The classical dynamics follows the differential equation, 
	
	\begin{equation} \label{2.2}
		\ddot{x} = -\omega^2 x - \lambda x^3
	\end{equation}
	
	which has an integral of motion $E = \frac{1}{2}\dot{x}^2 + \frac{1}{2} \omega^2 x^2 + \frac{\lambda}{4} x^4$.
	The motion is bounded between $-a \leq x \leq a$, where $\frac{1}{2}\omega^2 a^2 + \frac{\lambda}{4}a^4 = E$ and the exact time period $T$, of the motion is obtained from,
	
	\begin{equation} \label{2.3}
	T = 2 \int_{-a}^a \frac{dx}{\sqrt{\omega^2(a^2 - x^2) + \frac{\lambda}{2}(a^4 - x^4)}}
	\end{equation}
	
	For small $\lambda$ i.e. $\frac{\lambda a^2}{\omega^2}$ or $\frac{\lambda E}{\omega^4} << 1,$ we have
	
	\begin{equation} \label{2.4}
		T = \frac{2\pi}{\omega}[1 - \frac{3}{4\omega^4}\lambda E + ...],
	\end{equation}
	
	while for $\frac{\lambda E}{\omega^4} >> 1$, we have the asymptotic answer,
	
	\begin{align} \label{2.5}
		T &= 4\int_0^a \frac{dx}{\sqrt{2E - \frac{\lambda}{2}x^4}} \nonumber \\
		  &= \sqrt{\frac{2}{\lambda a^2}}\beta\Big(\frac{1}{2},\frac{1}{4}\Big) \nonumber  \\
		  &= \frac{1}{(\lambda E)^\frac{1}{4}} \frac{\Gamma(\frac{1}{2})\Gamma(\frac{1}{4})}{\Gamma(\frac{3}{4})} 
	\end{align}		
	
	In terms of the frequency $\Omega = \frac{2\pi}{T}$, we have for $\frac{\lambda E}{\omega^4} << 1$,
	
	\begin{equation} \label{2.6}
		\Omega^2 = \omega^2[1 + \frac{3}{2} \frac{\lambda E}{\omega^4} + ...]
	\end{equation}
	
	and for $\frac{\lambda E}{\omega^4} >> 1$, we have,
	
	\begin{equation} \label{2.7}
		\Omega^2 = \Big[2\pi \frac{(\lambda E)^\frac{1}{4} \Gamma(\frac{3}{4})}{\Gamma(\frac{1}{2})\Gamma(\frac{1}{4})} \Big]^2 \simeq 1.45 (\lambda E)^\frac{1}{2}
	\end{equation}
	
	A simple but reasonable approximation for $\Omega^2$ that covers the entire range of $\frac{\lambda E}{\omega^4}$ values is,
	
	\begin{equation} \label{2.8}
		\Omega^2 = \omega^2\Big[1 + 2\frac{\lambda E}{\omega^4}\Big]^\frac{1}{2}
	\end{equation}
	
	Turning to the quantum dynamics, while the exact solution can be obtained for the simple harmonic oscillation, very little is known about the dynamics of the anharmonic oscillator i.e. the one having the Hamiltonian,
	
	\begin{equation} \label{2.9}
		H = \frac{p^2}{2} + \frac{\lambda}{4}x^4
	\end{equation}
	
	where we have dropped the simple harmonic part of the potential in Eq.(\ref{2.1}) to concentrate only on the part where exact results are not available. The standard prescription is clear. The following steps need to be executed.
	
	\vspace{.1in}	
	
	\rm{i}) Find the spectrum of the Hermitian operator $H$, i.e. obtain all $E_n$ such that,
	
	\begin{equation} \label{2.10}
		H|u_n\rangle = E_n|u_n\rangle
	\end{equation}			
	
	In the coordinate representation each of the above states corresponds to the wavefunction $u_n(x)$.
	
	\vspace{.1in}	
	
	\rm{ii}) Any initial state $\psi(x)$ that is prepared can be expanded as,
	
	\begin{equation} \label{2.11}
		\psi_o(x) = \Sigma c_n u_n(x)
	\end{equation} 
	 
	 and the $c_n$ can be found as $c_n = \int_{-\infty}^\infty \psi_o(x) u_n^*(x) dx$.
	 
	\vspace{.1in}	 
	 
	 \rm{iii}) Each $u_n(x)$ time develops as $u_n(x) e^{-\frac{\imath E_n t}{\hbar}}$ and hence at time $t$
	 
	 \begin{equation} \label{2.12}
	 	\psi(x,t) = \Sigma c_n u_n(x) e^{-\frac{\imath E_n t}{\hbar}}
	 \end{equation}
	
	\vspace{.1in}	
	
	Performing the above sum gives the exact final answer. Since no closed form expression of $E_n$ is known(not even the ground state), only a numerical evaluation is possible. This is where an analysis of the time dependence of the moments of the probability distribution can give closed form answers which can be of use. The arbitrary initial state has moments $\langle x\rangle$, $\langle x^2\rangle$ ... $\langle x^n\rangle$ etc which are known and we want to write down the dynamics of the first few moments. Starting with the operator $x$, we have using the equation of motion of Eq.(\ref{1.4})
	
	\begin{subequations} \label{2.13}
	\begin{equation} 
		\frac{d}{dt}\langle x \rangle = \langle p \rangle
	\end{equation}
	\begin{equation}
		\frac{d}{dt}\langle p \rangle = -\lambda \langle x^3 \rangle
	\end{equation}	
	\end{subequations}		 
	 
	 and hence,
	 
	 \begin{equation}	\label{2.14}
	 	\frac{d^2}{dt^2} \langle x \rangle = -\lambda \langle x^3 \rangle
	 \end{equation}
	 
	The problem is immediately clear - the first moment has coupled to the third moment. We will see that this is a recurring feature -  the dynamics of a lower moment brings in higher moments, making the problem unsolvable. We note that,
	
	\begin{equation} \label{2.15}
		\langle x^3 \rangle = \langle x \rangle^3 + 3 V \langle x \rangle + S
	\end{equation}
	
	where 
	
	\begin{align} \label{2.16}
		V &= \langle x^2 \rangle - \langle x \rangle^2 \\
		  &= \langle(x - \langle x \rangle)^2 \rangle \nonumber
	\end{align}
	
	is the variance of the quantum state and 
	
	\begin{equation} \label{2.17}
		S = \langle (x - \langle x \rangle)^3 \rangle
	\end{equation}
	
	is its skewness. The variance and the skewness are quintessential quantum features and they affect the dynamics of the mean position which can be written as 
	
	\begin{equation} \label{2.18}
		\frac{d^2}{dt^2}\langle x \rangle = -3\lambda V \langle x \rangle - \lambda \langle x \rangle^3 - \lambda S
	\end{equation}
	
	Unlike the classical case, we cannot draw any conclusion about the dynamics of $\langle x \rangle$ unless we have information about the dynamics of V and S. Consequently we proceed to find the dynamics of $V$ by writing (using Eq.(\ref{1.4}) with $O = x^2$),
	
	\begin{subequations} \label{2.19}
	\begin{equation}
		\frac{d}{dt}\langle x^2 \rangle = \langle xp + px \rangle
	\end{equation}
	\begin{align}
		\frac{d^2}{dt^2}\langle x^2 \rangle &= \frac{d}{dt}\langle xp + px \rangle  \nonumber \\
											&= 2\langle p^2 \rangle - 2\lambda \langle x^4 \rangle 
	\end{align}
	\end{subequations}
	
	The parameter in the problem is once again the total energy $e$ which is the expectation value of the Hamiltonian(a constant of motion)
	
	\begin{equation} \label{2.20}
		e = \frac{\langle p^2 \rangle}{2} + \frac{\lambda}{4} \langle x^4 \rangle
	\end{equation}
	
	Substituting for $\langle p^2 \rangle$ in Eq.(\ref{2.19}) from Eq.(\ref{2.20}), we get,
	
	\begin{equation} \label{2.21}
		\frac{d^2}{dt^2} \langle x^2 \rangle = 4e - 3\lambda \langle x^4 \rangle
	\end{equation}
	
	We note that
	
	\begin{equation} \label{2.22}
		\langle x^4 \rangle = K + 4S\langle x \rangle + 6V\langle x \rangle^2 + \langle x \rangle^4
	\end{equation}
	
	where $K$ is the kurtosis $\langle (x - \langle x \rangle)^4 \rangle$ of the distribution.
	
	Further, 
	
	\begin{align} \label{2.23}
		\frac{d^2}{dt^2}\langle x \rangle^2 &= 2\frac{d}{dt}[\langle x \rangle \langle p \rangle] \nonumber \\
											&= 2\langle p \rangle^2 - 6\lambda V \langle x \rangle^2 - 2\lambda \langle x \rangle^4 - 2\lambda S \langle x \rangle
	\end{align}

	Writing $\langle x^2 \rangle = V + \langle x \rangle^2$ in Eq. (\ref{2.21}) and using Eqs. (\ref{2.22}) and (\ref{2.23}), we get,
	
	\begin{align} \label{2.24}
		\frac{d^2}{dt^2}V = 4e \ - \ &3\lambda K - 12\lambda V \langle x \rangle^2 - 10\lambda S \langle x \rangle \nonumber \\ &- \lambda \langle x \rangle^4 - 2\langle p \rangle^2
	\end{align}
	
	We note that the generic feature - the second moment(variance) couples to the skewness and the kurtosis.
	
	\vspace{.1in}
	
	We need to write down the dynamics of $S$ and $K$. Following an identical procedure as we did for the dynamics of $V$, we obtain for $\langle x^3 \rangle$(the new part of $S$),
	
	\begin{equation} \label{2.25}
		\frac{d^4}{dt^4}\langle x^3 \rangle = -\frac{63}{10}\lambda \frac{d^2}{dt^2}\langle x^5 \rangle + 9\lambda\hbar^2 \langle x \rangle - \frac{9}{2}\lambda^2 \langle x^7 \rangle
	\end{equation}
	
	As one can see, unless a truncation scheme is used, this procedure will get out of hand. The simplest truncation is a following approximation $K = \langle(x - \langle x \rangle)^4\rangle \propto \langle (x - \langle x \rangle)^2\rangle^2$. The proportionality constant can be pre-assigned(we can take it to be $3$ if we consider a Gaussian approximation) or we can write it as an unknown parameter $\beta$ and fix it self consistently, i.e. we can write $K = 3\beta V^2$ and determine $\beta$ using a secondary argument. As for $S$, we cannot have any factoring form for it. A purely phenomenological approach would be to set $S = \beta_1\langle x \rangle + \beta_2\langle x \rangle^3 + ...$ A more systematic approach is to consider the next higher moment $\langle (x - \langle x \rangle)^5\rangle$ and note that this can factor as VS and so on, while $\langle(x - \langle x \rangle)^7\rangle$ factors as $V\langle(x - \langle x \rangle)^5\rangle$, $V^2S$ or $KS$. We then have
	
	\begin{subequations} \label{2.26}
	\begin{equation}
		K = 3\beta V^2
	\end{equation}
	\begin{equation}
		\langle (x - \langle x \rangle)^5\rangle = \gamma VS
	\end{equation}
	\begin{equation}
	\langle (x - \langle x \rangle)^7\rangle = \delta V^2S
	\end{equation}
	\end{subequations}
	
	and with this Eqs. (\ref{2.9}),(\ref{2.15}),(\ref{2.16}),(\ref{2.24}) and (\ref{2.25}) are closed.
	
	\vspace{.1in}
			
	To demonstrate the utility of this approach we will begin with the approximation that $S = 0$ i.e. an initially symmetric(about the centre) wave packet will remain almost symmetric as it moves in this monotonic symmetric potential. We will also set $\beta = 1$ i.e. assume that an initially Gaussian packet will remain approximately Gaussian. Having done this, we have the coupled dynamical system,
	
	\begin{subequations} \label{2.27}
	\begin{equation}
		\frac{d^2}{dt^2} \langle x \rangle = -3\lambda V \langle x \rangle - \lambda \langle x \rangle^3
	\end{equation}
	\begin{equation}
		\frac{d^2}{dt^2}V = 4e - 9\lambda V^2 - 12\lambda V \langle x \rangle^2 - \lambda \langle x \rangle^4 - 2\langle p \rangle^2
	\end{equation}
	\end{subequations}
	
	This coupled set has only one fixed point
	
	\begin{align} \label{2.28}
		\langle x \rangle^* &= \langle p \rangle^* = 0 \nonumber \\
						V^* &= \Big(\frac{4e}{9\lambda}\Big)^\frac{1}{2}
	\end{align}
	
	To test the stability of the fixed point, we carry out a linear stability analysis and find the perturbation $\delta x$ and 
	$\delta V$ satisfies,
	
	\begin{subequations} \label{2.29}
	\begin{equation}
		\frac{d^2}{dt^2}\delta x = -3\lambda V^*\delta x
	\end{equation}
	\begin{equation}
		\frac{d^2}{dt^2}\delta V = -18\lambda V^*\delta V
	\end{equation}
	\end{subequations}
	
	Clearly the fixed point is a centre with the frequency of oscillation for $\delta x$ given by,
	
	\begin{align} \label{2.30}
		\Omega^2 &= 3\lambda V^* \nonumber \\
				 &= 2 (e\lambda)^\frac{1}{2}
	\end{align}
	
\begin{figure}[h]
	\begin{subfigure}{.5\textwidth}
		\includegraphics[width=\linewidth]{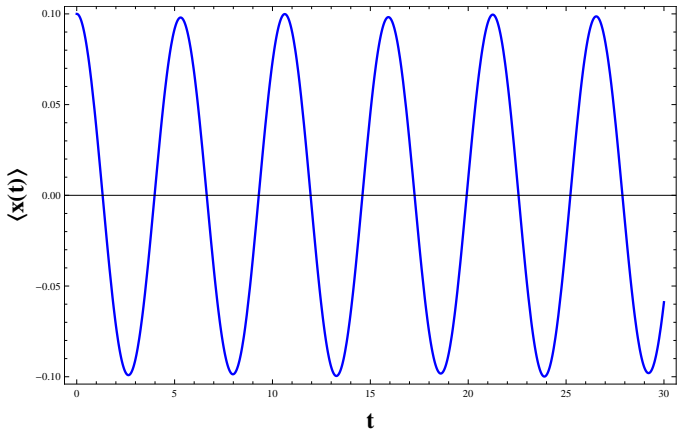}	
		\caption{With $\langle x(0) \rangle = .1$, $\langle \dot{x}(0) \rangle = 0$, $V(0) = V_o = .471405$ and $\dot{V}(0) = 0$. $\langle x \rangle$ oscillates about zero with a frequency $\Omega = 1.185$ as predicted by the expression $\sqrt{3\lambda V_o}$ from Eq.(\ref{2.30})}
		\label{Fig 2a}
	\end{subfigure}	
	
	\begin{subfigure}{.5\textwidth}
		\includegraphics[width=\linewidth]{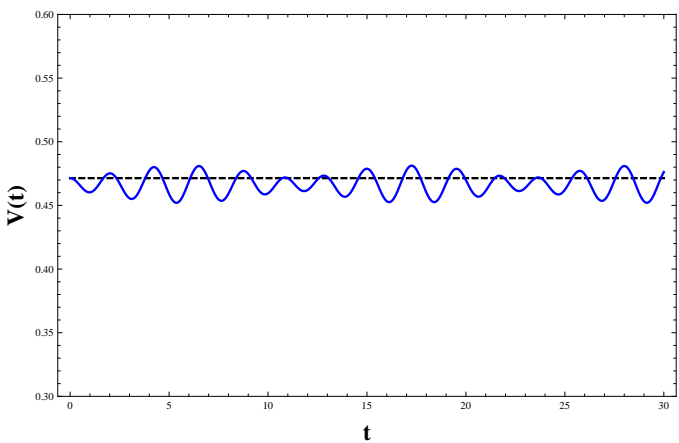}	
		\caption{With $V(0) = V_o = .471405$ and $\dot{V}(0) = 0$. The variance $V(t)$ oscillates about its fixed point(black dashed line) $V_o = \frac{2}{3}\sqrt{\frac{e}{\lambda}}$ from Eq.(\ref{2.28})}
		\label{Fig 2b}
	\end{subfigure}	
	\caption{Numerics of $\langle x \rangle$ and $V(t)$ vs $t$ for $e = .5$, $\lambda = 1$.}
	\label{Fig. 2}
	\end{figure}

	Hence the average position of the wave packet will oscillate with the frequency $\Omega$ in the linearised approximation. As for $\delta V$ it will oscillate with the frequency $\Omega^\prime$, where $\Omega^{\prime2} = 12(e\lambda)^{\frac{1}{2}}$. At this order the solution for the variance can be written as
	
	\begin{equation} \label{2.31}
		V = \frac{2}{3}\Big(\frac{e}{\lambda}\Big)^{\frac{1}{2}} + A \cos{\Omega^\prime t} + B \sin{\Omega^\prime t}
	\end{equation}
	
	where $A$ and $B$ are constants to be determined from initial conditions. An initial value of $\frac{dV}{dt} = 0$ for the derivative of $V$ makes the constant $B = 0$ and initial value of $V_o = \frac{2}{3}\Big(\frac{e}{\lambda}\Big)^\frac{1}{2}$ makes $A = 0$. Consequently if the initial packet is given the variance $\frac{2}{3}\Big(\frac{e}{\lambda}\Big)^\frac{1}{2}$, then the dynamics of the variance should remain almost unchanged as the system evolves with time. The numerical solutions of Eqs.(\ref{2.27}) has been shown in Fig. (\ref{Fig. 2}). 
	
	\vspace{.1in}

	If we prescribe initially a Gaussian packet
	 
	\begin{equation} \label{2.32}
		\psi_o(x) = \frac{1}{\pi^\frac{1}{4}\Delta_o^\frac{1}{2}} e^{-\frac{(x - a)^2}{2\Delta_o^2}}
	\end{equation}	 
	
	then the energy is a constant of motion works out as ($e$ is scaled by $\hbar^\frac{4}{3}$ and length is scaled by $\hbar^\frac{1}{3}\lambda^{-\frac{1}{6}}$)
	
	\begin{align} \label{2.33}
		e &= \frac{\hbar^2}{4\Delta_o^2} + \frac{\lambda}{4}[a^4 + 3a^2\Delta_o^2 + \frac{3}{4}\Delta_o^4] \nonumber \\
		  &= \frac{1}{8\overline{V_o}} + \frac{1}{4}(3\overline{V_o}^2 + 6a^2\overline{V_o} + a^4)
	\end{align}
	
	For small `$a$', we have (using Eq.(\ref{2.28}))
	
	\begin{equation} \label{2.34}
		\frac{9}{4}\overline{V_o}^2 \simeq \frac{1}{8\overline{V_o}} + \frac{3}{4}\overline{V_o^2}
	\end{equation}		
	
	leading to $\overline{V_o}^3 = \frac{1}{12}$ or $\overline{V_o} \simeq .083$. Hence we predict that if a Gaussian packet $\frac{1}{\pi^\frac{1}{4}\Delta_o^\frac{1}{2}} e^{-\frac{(x - a)^2}{2\Delta_o^2}}$ with $\Delta_o^2 = 2\overline{V_o}$ and $a << 1$ is allowed to time develop with the Hamiltonian of Eq.(\ref{2.9}) then it will be approximately shape invariant in time with the centre oscillating with frequency $\sqrt{3\lambda V_o}$.
	
	\vspace{.1in}	
	
	We can ask how reliable our results, based on a Gaussian approximation, are. In the high energy limit one should end up with the classical dynamics and hence the frequency should match the classical result shown in Eq. (\ref{2.7}). While the scaling behaviour agrees exactly, there is discrepancy in the pre-factor. The frequency $\Omega$ from our Eq. (\ref{2.30}) is $\Omega \simeq \sqrt{2}(e\lambda)^\frac{1}{4}$, while in the classical limit it is $1.2(e\lambda)^\frac{1}{4}$ as can be seen from Eq. (\ref{2.7}). This means that we need $\beta < 1$ in Eq. (\ref{2.26}) which is consistent with the nature of the potential in that the probability distribution should have a smaller tail than a Gaussian.   
	
	\section{The Volcano Potential: escape over a barrier}
	
\begin{figure}[h]
		\includegraphics[width=\linewidth]{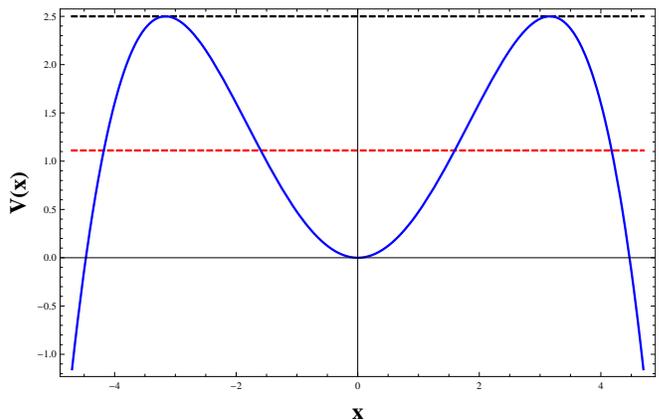}	
			\caption{(color online) Volcano Potential with $\omega = 1$, $\lambda = 0.1$. The black dashed line shows the height of the potential $\frac{\omega^4}{4\lambda}$, the red dashed line shows the critical energy $\frac{\omega^4}{9\lambda}$ (due to the Gaussian approximation) below which oscillation about zero is observed.}
			\label{Fig. 3}
	\end{figure}

	The potential here is volcano shaped \cite{30} as shown in Fig. (\ref{Fig. 3}). We now have 
	
	\begin{equation} \label{3.1}
		V(x) = \frac{1}{2}\omega^2 x^2 - \frac{\lambda}{4}x^4
	\end{equation}
	
	The classical dynamics corresponds to oscillation in the well for all energies less than $\frac{\omega^4}{4\lambda}$. The particle escapes from the well for $E > \frac{\omega^4}{4\lambda}$.
	
	\vspace{.1in}	
	
	If we are to look at the quantum mechanical problem, then the mean $\langle x \rangle$ and the variance $V$ satisfies(adding the contribution of the quadratic term to Eq. (\ref{2.18}) and Eq.(\ref{2.24})
	
	\begin{subequations} \label{3.2}
	\begin{equation}
		\frac{d^2}{dt^2}\langle x \rangle = -(\omega^2 - 3\lambda V)\langle x \rangle + \lambda \langle x \rangle^3 + \lambda S
	\end{equation}
	\begin{align}
		\frac{d^2}{dt^2}V = 4e &- 4\omega^2 V + 9\lambda V^2 \nonumber \\ &- 2\omega^2 \langle x \rangle^2 + 12\lambda V \langle x \rangle^2 +\lambda \langle x \rangle^4 \nonumber \\ &- 2\langle p \rangle^2 + 10\lambda S \langle x \rangle
	\end{align}
	\end{subequations}
	
	In the simplest approximation we set $S = 0$ i.e. ignore the skewness. The only relevant fixed point as before is $x^* = 0$, $V^* = V_o$ satisfying
	
	\begin{equation} \label{3.3}
		9\lambda V^2 - 4\omega^2 V + 4e = 0
	\end{equation}
	
	leading to
	
	\begin{equation} \label{3.4}
		V_o = \frac{2\omega^2}{9\lambda} \Big(1 \pm \sqrt{1 - \frac{9\lambda e}{\omega^4}}\Big)
	\end{equation}
	
	Existence of the fixed point requires $e < \frac{\omega^4}{9\lambda}$. The stability of the fixed point $x^* = 0$ and $V^* = V_o$ is found by linearising around it. This gives, 
	
	\begin{subequations} \label{3.5}
	\begin{equation}
		\frac{d^2}{dt^2}\langle x \rangle = -(\omega^2 - 3\lambda V_o)\delta x
	\end{equation}
	\begin{equation}
		\frac{d^2}{dt^2}V = -(4\omega^2 - 18\lambda V_o)\delta V
	\end{equation}
	\end{subequations}
	
\begin{figure}
	\begin{subfigure}{.5\textwidth}
		\includegraphics[width=\linewidth]{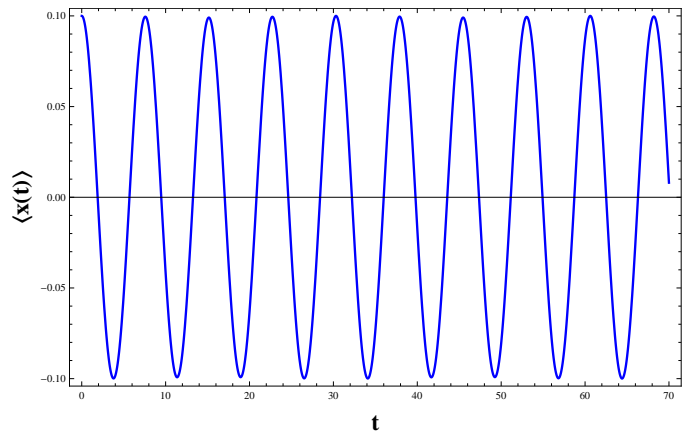}	
		\caption{With $\langle x(0) \rangle = .1$ and $\langle \dot{x}(0) \rangle = 0$. $\langle x \rangle$ oscillates about zero.}
		\label{Fig. 4a}
	\end{subfigure}	
	
	\begin{subfigure}{.5\textwidth}
		\includegraphics[width=\linewidth]{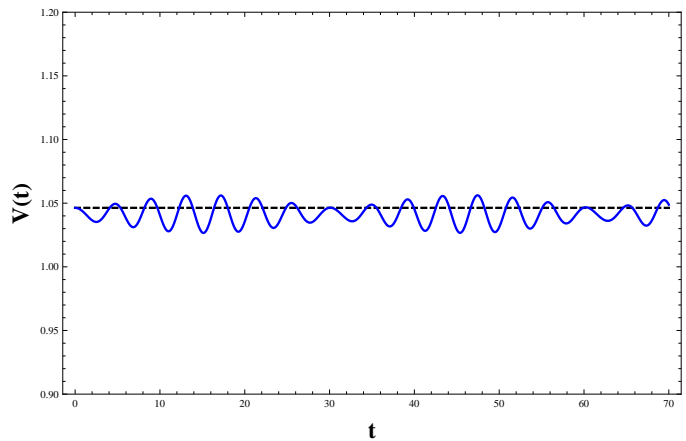}	
		\caption{With $V(0) = V_o = 1.04633$ (fixed point(black dashed line) for $V$ given by Eq. (\ref{3.4})) and $\dot{V}(0) = 0$}.
		\label{Fig. 4b}
	\end{subfigure}	
	\caption{Numerics of $\langle x \rangle$ and $V(t)$ vs $t$ for $e = .8$, $\omega = 1$ and $\lambda = 0.1$.}
	\label{Fig. 4}
	\end{figure}

	The fixed point $V^*$ is stable if $\omega^2 > \frac{9}{2}\lambda V_o$. This is always satisfied for the negative sign of $V_o$ in Eq.(\ref{3.4}) which is the physically consistent sign. The negative sign is physically relevant because it shows that $\frac{\lambda e}{\omega^4} << 1$, one has $V_o \simeq \frac{e}{\omega^2}$ which is consistent with the dynamics in a single well. Hence the origin is stable for all energies $e < \frac{\omega^4}{9 \lambda}$ i.e. for approximately the lower half of the well. Numerical plots of $\langle x \rangle$ and $V$ for $e$ less than this critical value has been shows in Fig.(\ref{Fig. 4}).
	
	\vspace{.1in}	
	
\begin{figure}
	\begin{subfigure}{.5\textwidth}
		\includegraphics[width=\linewidth]{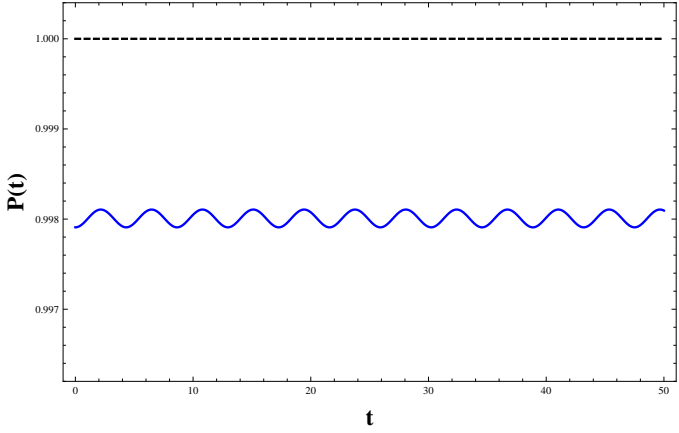}	
		\caption{Figure shows the probability $P(t)$, of oscillation about zero without escaping with $e = .8$ and for values used in Fig. (\ref{Fig. 4}).}
		\label{Fig. 5a}
	\end{subfigure}	
	
	\begin{subfigure}{.5\textwidth}
		\includegraphics[width=\linewidth]{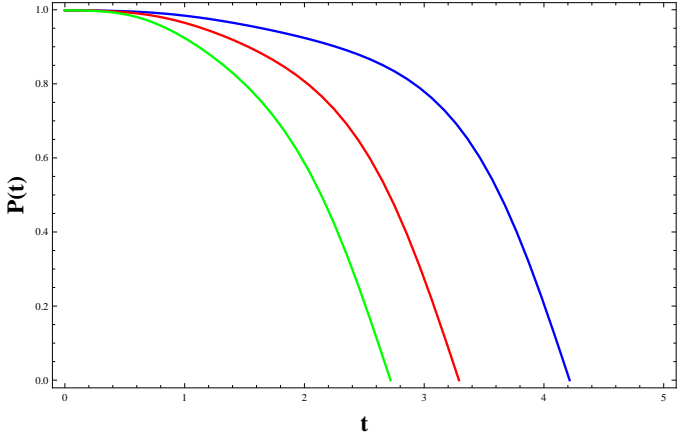}	
		\caption{(color online) Figure shows $P(t)$ vs $t$ for $\langle x(0) \rangle = 0.1$, $V(0) = 1.0$ and various values of energy, $e > \frac{\omega^4}{9\lambda}$. The blue, red and green curves corresponds to energy values of $1.2, 1.5$ and $2.0$. For increasing energy values, the time,$T$ taken for the dynamics to escape the well decreases.}
		\label{Fig. 5b}
	\end{subfigure}	
	\caption{Numerics of the probability $P(t)$ vs $t$ with $\omega = 1$ and $\lambda = .1$.}
	\label{Fig. 5}
	\end{figure}

\begin{figure}
		\includegraphics[width=\linewidth]{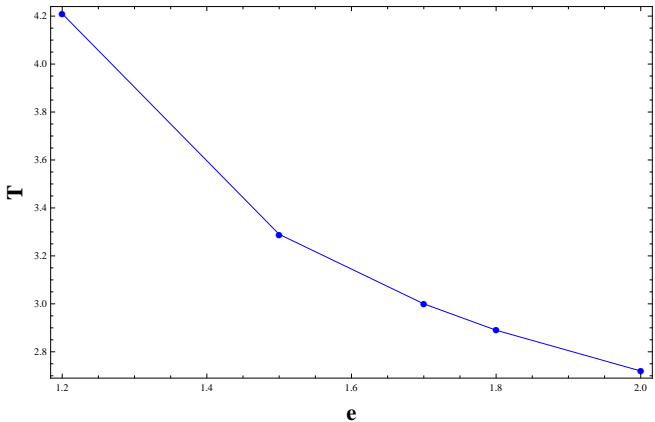}	
		\caption{Time $T$ vs $e$ for which the dynamics escapes the well for various values of $e > \frac{\omega^4}{9\lambda}$.}
		\label{Fig. 6}
	\end{figure}

	 For energies in the upper half of the well where the classical motion is simply an oscillation about the centre of the well, the quantum fluctuations cause the particle to escape from the well. At any time `$t$' we can define the probability of the particle being in the well as $P(t) = \frac{1}{\sqrt{2\pi V}}\int_{-\frac{\omega}{\sqrt{\lambda}}}^{\frac{\omega}{\sqrt{\lambda}}} e^{-\frac{(x - \langle x \rangle(t))^2}{2V(t)}} dx$. For different values of the total energy(found by evaluating the expectation values of the initial packet) 
	
	\begin{equation} \label{3.6}
		E = \frac{\langle p^2 \rangle_o}{2} + \frac{\omega^2}{2}\langle x^2 \rangle_o - \frac{\lambda}{4}\langle x \rangle_o^4
	\end{equation}
	
	we show $P(t)$ vs $t$ in Fig. (\ref{Fig. 5}). Fig. (\ref{Fig. 6}) shows time $T$ vs energy $e$ for which the dynamics escapes the well. It is found that as energy increases the time taken for the dynamics of $\langle x \rangle$ to escape the well decreases as expected.

	\vspace{.1in}	
	
	As is apparent from the figure, the escape occurs only for $E > \frac{\omega^4}{9\lambda}$. Quantum fluctuations would actually cause all initial states with $E > 0$ to eventually escape. The restriction on energy found above is, once again(as in the last section), a result of the Gaussian approximation. The Gaussian approximation shows up in two different places - once in setting $S = 0$ and then again in writing the kurtosis $K = 3V^2$. We can introduce the non dimensional parameter $\beta$ shown in Eq. (\ref{2.26}) and self consistently determine $\beta$ by requiring the escape occurring for as low an energy as possible. 
	
	\section{The Double Well Potential}
	
\begin{figure}[h]
		\includegraphics[width=\linewidth]{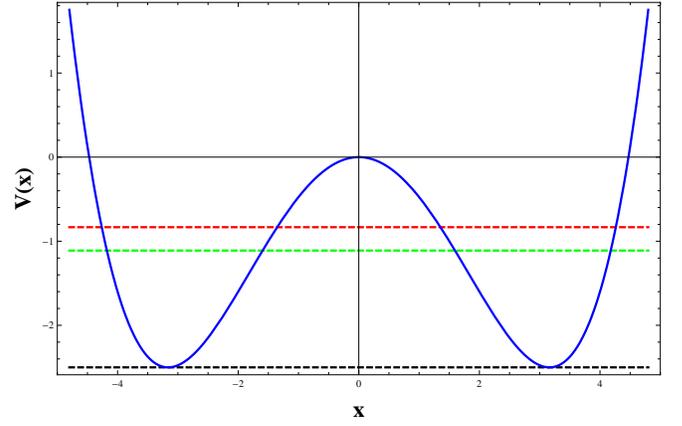}	
		\caption{(color online) Double well potential. Black dashed line shows the bottom of the well $\frac{\omega^4}{4\lambda}$, green line is the minimum energy $\frac{\omega^4}{9\lambda}$, for which the variance $V$ exists. The red line is the minimum energy $\frac{\omega^4}{12\lambda}$ needed to stabilize $\langle x \rangle$ about zero.}
		\label{Fig. 7}
	\end{figure}

	In this section we consider the double well potential given by(shown in Fig. \ref{Fig. 7}) 
	
	\begin{equation} \label{4.1}
		V(x) = -\frac{1}{2}\omega^2 x^2 + \frac{\lambda}{4}x^4
	\end{equation}
	
	The classical dynamics is given by 
	
	\begin{equation} \label{4.2}
		\frac{d^2 x}{dt^2} = \omega^2 x - \lambda x^3
	\end{equation}
	
	The fixed points are $x^* = 0$ and $x^* = \pm \sqrt{\frac{\omega^2}{\lambda}}$. The fixed point $x^* = 0$ is clearly unstable for $\omega > 0$. We write the total energy of the system(conserved quantity) as 
	
	\begin{equation} \label{4.3}
		E = \frac{p_o^2}{2} - \frac{1}{2}\omega^2 x_o^2 + \frac{\lambda}{4}x_o^4
	\end{equation}	 
	
	where $x_o$ is the initial position and $p_o$ is the initial momentum of the system. Without loss of generality we can consider initial conditions with $p_o = 0$. It is easy to see that for $E < 0$(refer Fig. \ref{Fig. 7}), the motion is confined to either the left well or the right well depending on whether the $x_o$ was in the vicinity of $\sqrt{\frac{\omega^2}{\lambda}}$ or $-\sqrt{\frac{\omega^2}{\lambda}}$. For $E > 0$(refer Fig. \ref{Fig. 7}) the motion takes place around $x^* = 0$ spanning both the wells.
	
	\vspace{.1in}	
	
	For the quantum problem we need to use the corresponding equations for the moments as found in Sec. $\rm{III}$ with the sign of $\omega^2$ and $\lambda$ reversed. For the mean position of the wave packet we use Eq. (\ref{3.2}) to write the coupled dynamical systems(we are in the Gaussian approximation with $S = 0$)
	
	\begin{equation} \label{4.4}
		\frac{d^2}{dt^2}\langle x \rangle = (\omega^2 - 3\lambda V)\langle x \rangle - \lambda \langle x \rangle^3
	\end{equation}
	
	\begin{align} \label{4.5}
		\frac{d^2}{dt^2}V = 4e &+ 4\omega^2 V - 9\lambda V^2 \nonumber \\ &+ 2\omega^2 \langle x \rangle^2 - 12\lambda V \langle x \rangle^2 \nonumber \\ &- \lambda \langle x \rangle^4 - 2\langle p \rangle^2
	\end{align}
	
	We look for fixed points with $\langle p \rangle = 0$ and $W = \frac{dV}{dt} = 0$. The values of $\langle x \rangle$ and $V$ at the fixed point are $\langle x \rangle = a$ and $V = V^*$. There are three pairs
	
	\vspace{.1in}	
	
	$A)$ $a = 0$ and $V = V_o$ with 
	
	\begin{equation} \label{4.6}
		4e + 4\omega^2 V_o - 9\lambda V_o^2 = 0	
	\end{equation}		 
	
	\vspace{.1in}	
	
	$B)$ $a^2 = \frac{\omega^2 - 3\lambda \overline{V_o}}{\lambda}$, $V = \overline{V_o}$ with 
	
	\begin{equation} \label{4.7}
		4e + \frac{\omega^4}{\lambda} - 8\omega^2 \overline{V_o} + 18\lambda \overline{V_o}^2 = 0
	\end{equation}
		
	Our focus will be on the fixed point $A$ which has $a = 0$. This fixed point is always unstable in the classical case for $E < 0$. The issue is, will quantum fluctuations stabilize it in the region where it is classically unstable. Then the fixed point $A$ needs real values for two variables $`a'$ and $V_o$. For $`a'$ it is zero while for $V_o$ it is(from Eq. (\ref{4.6}))
	
	\begin{equation} \label{4.8}
		V_o = \frac{2\omega^2}{9\lambda}\Big[1 \pm \sqrt{1 + \frac{9\lambda e}{\omega^4}}\Big]
	\end{equation}
	
	Clearly a real $V_o$ exists if $e > -\frac{\omega^4}{9\lambda}$ i.e. if the energy resides approximately in the upper half of the well shown in Fig. \ref{Fig. 7}. With the existence issue settled, we need the stability. Stability of $V_o$ follows after Eq. (\ref{4.5}) is linearised around the fixed point $(0,V_o)$. This leads to 
	
	\begin{equation} \label{4.9}
		\frac{d^2}{dt^2} \delta V = (4\omega^2 - 18\lambda V_o)\delta V
	\end{equation}
	
	For stability(i.e. oscillations) we need $4\omega^2 < 18\lambda V_o$ and hence the $V_o$ that we need is
	
	\begin{equation} \label{4.10}
		V_o = \frac{2\omega^2}{9\lambda}\Big[1 + \sqrt{1 + \frac{9\lambda e}{\omega^4}}\Big]
	\end{equation}	 
	
	But we still need the stability of the fixed point $a = 0$. Linearising Eq. (\ref{4.4}) we see that this fixed point will be stable if 
	
	\begin{equation}	\label{4.11}
		3\lambda V_o > \omega^2
	\end{equation}
	
	From Eq. (\ref{4.10}) this implies $\frac{2}{3}(1 + \sqrt{1 + \frac{9\lambda e}{\omega^4}}) > 1$ or 
	
	\begin{equation} \label{4.12}
		e > -\frac{\omega^4}{12\lambda}
	\end{equation}
	
\begin{figure}
	\begin{subfigure}{.5\textwidth}
		\includegraphics[width=\linewidth]{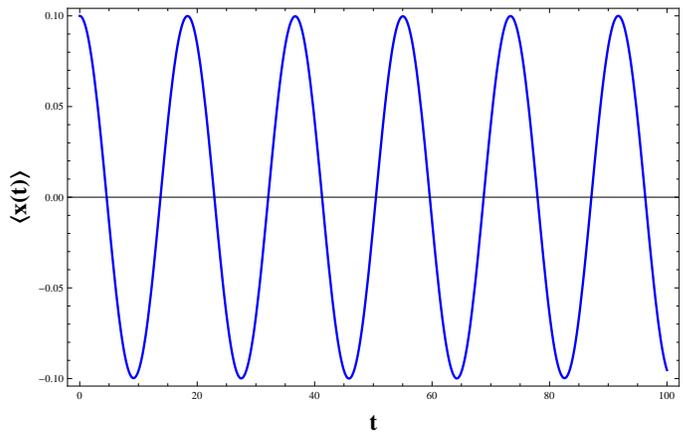}	
		\caption{Plot of $\langle x \rangle$ vs $t$ with $\langle x(0) \rangle = 0.1$, $\langle \dot{x}(0) \rangle = 0$ and $V(0) = V_o = 3.72941$(fixed point $V_o$ given by Eq. (\ref{4.10})) and $\dot{V}(0) = 0$.}
		\label{Fig. 8a}
	\end{subfigure}	
	
	\begin{subfigure}{.5\textwidth}
		\includegraphics[width=\linewidth]{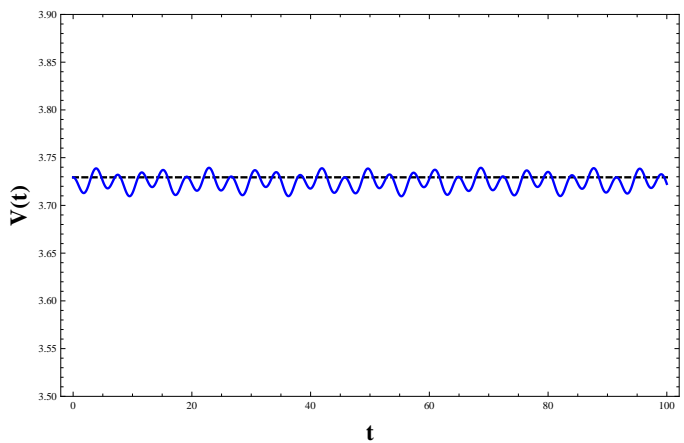}	
		\caption{Plot of $V(t)$ vs $t$ with $V(0) = V_o = 3.72941$ and $\dot{V}(t) = 0$.}
		\label{Fig. 8b}
	\end{subfigure}	
	\caption{Numerics of $\langle x \rangle$ and $V(t)$ with $\omega = 1$, $\lambda = 0.1$ and $e = -0.6$.}
	\label{Fig. 8}
	\end{figure}

	The numerics for $\langle x \rangle$ and $V$ corresponding to the fixed point A, i.e. $(0, V_o)$, where $V_o$ is given by Eq. (\ref{4.10}) is shown in Fig. \ref{Fig. 8}.
	
	\vspace{.1in}	
	
	Hence in the Gaussian approximation all states with mean energy $e > e_g = -\frac{\omega^4}{12\lambda}$(shown in Fig. \ref{Fig. 7}) will escape from the well and oscillate between the two wells which is the phenomenon of tunnelling. Quantum mechanical tunnelling occurs not just from the top one third of the well but from a far bigger region. As is shown in all quantum mechanics text books, the lower energy levels of the double well oscillation come in the form of closely spaced doublets. The ground state(symmetric wave function with peaks in the two wells and no nodes) is separated by a very small amount from the first excited state(the antisymmetry version of the ground state). A linear combination of these two states can be localised in one of the wells and for such an initial state the wave function time develops in a manner such that after a time $\tau \simeq \frac{\hbar}{\Delta E}$($\Delta E$ is the energy difference between ground and first excited states) it is localised in the other well. Around what energy is the doublet centred? It is easy to see that each well is locally a simple harmonic oscillator with the Hamiltonian $H = \frac{p^2}{2m} + m\omega^2 x^2$ and hence the relevant energy is $\frac{1}{2}\sqrt{2}\omega \hbar$. Thus for an energy $E$ as low as $-\frac{\omega^4}{4\lambda} + \frac{\hbar \omega}{\sqrt{2}}$, the tunnelling can occur. The question is how does one take this into account.
	
	The primary error, as can be easily guessed, is in the Gaussian approximation. We can remove this phenomenologically by including the skewness in the form $S = \beta_1\langle x \rangle + \beta_2\langle x \rangle^3$, where $\beta_1$ and $\beta_2$ are adjustable parameters. This changes Eq. (\ref{4.4}) to 
	
	\begin{equation} \label{4.13}
		\frac{d^2}{dt^2} \langle x \rangle = (\omega^2 - \beta_1 - 3\lambda V)\langle x \rangle - (\lambda + \beta_2)\langle x \rangle^3
	\end{equation}
	
	At the end of this section we will show for what value of $\beta_1$, the origin becomes stable for the entire range of its existence. Alternatively, we can keep $S = 0$ and use $K = 3\beta V^2$. Repeating the calculation above it is easy to see that $\beta = \frac{1}{3}$ is necessary for tunnelling to occur from anywhere in the well.
	
	\vspace{.1in}	
	
	We now need to discuss the stability of the fixed point with $\langle x^* \rangle \neq 0$, i.e. the fixed point $B$. In this case the more natural energy variable is $\overline{e} = e - (-\frac{\omega^4}{4\lambda}) = e + \frac{\omega^4}{4\lambda}$ which is the energy relative to the bottom of the well. From Eq. (\ref{4.7}), we have 
	
	\begin{equation} \label{4.14}
		\overline{V_o} = \frac{2\omega^2}{9\lambda}\Big[1 \pm \sqrt{1 - \frac{9\lambda \overline{e}}{2\omega^4}}]\Big]
	\end{equation}
	
	Thus, this fixed point ceases to exist if $\overline{e} > \frac{2\omega^4}{9\lambda}$. As for stability of the fixed point, linearising about the fixed point we find
	
	\begin{subequations} \label{4.15}
	\begin{align}
		\frac{d^2}{dt^2}\delta x &= (\omega^2 - 3\lambda \overline{V_o} - 3\lambda a^2)\delta x \nonumber -3\lambda a \delta V \\ 								 &= -2(\omega^2 - 3\lambda \overline{V_o})\delta x - 3\lambda a \delta V
	\end{align}
	\begin{align}
		\frac{d^2}{dt^2}\delta V &= (4\omega^2 a - 24\lambda \overline{V_o} a - 4\lambda a^3)\delta x \nonumber \\ & \hspace{.2in}+ (4\omega^2 - 18\lambda \overline{V_o} -12\lambda a^2)\delta V \nonumber \\ &= -12\lambda \overline{V_o} a 	\delta x -(8\omega^2 - 18\lambda \overline{V_o})\delta V
	\end{align}
	\end{subequations}
	
	The stability matrix is 
	
	\begin{equation} \label{4.16}
		M = -\begin{pmatrix}
				2(\omega^2 - 3\lambda \overline{V_o}) & 3\lambda a \\
				12\lambda \overline{V_o} a & 8\omega^2 - 18\lambda \overline{V_o}
			\end{pmatrix}
	\end{equation}
	
	and for stability we require both eigenvalues to be positive. This implies that the trace has to be positive and the determinant has to be positive as well. Hence
	
	\begin{subequations} \label{4.17}
	\begin{equation}
		5\omega^2 - 12\lambda \overline{V_o} > 0
	\end{equation}
	\begin{equation}
		2\omega^4 - 15\omega^2 \lambda \overline{V_o} + 27\lambda^2 \overline{V_o}^2 > 0
	\end{equation}
	\end{subequations}
	
	The latter condition can be expressed as 
	
	\begin{equation} \label{4.18}
		(\omega^2 - 3\lambda \overline{V_o})(2\omega^2 - 9\lambda \overline{V_o}) > 0
	\end{equation}
	
	and hence the conditions of Eqs. (\ref{4.17}) needs to be simultaneously satisfied.
	
	For the variance the relevant fixed point(the correct low energy limit is attained) is
	
	\begin{equation} \label{4.19}
		\overline{V_o} = \frac{2\omega^2}{9\lambda} \Big[1 - \sqrt{1 - \frac{9\lambda \overline{e}}{2 \omega^4}} \Big]
	\end{equation}
	
\begin{figure}
	\begin{subfigure}{.5\textwidth}
		\includegraphics[width=\linewidth]{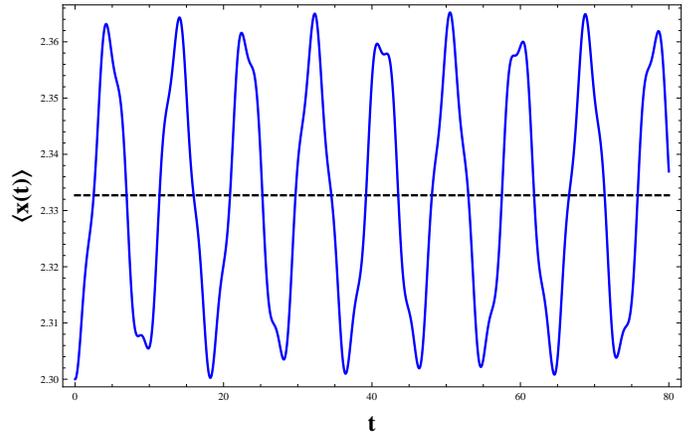}	
		\caption{Plot of $\langle x \rangle$ vs $t$ with $\langle x(0) \rangle = 2.3$, $\langle \dot{x}(0) \rangle = 0$ and $V(0) = V_o = 1.51949$(fixed point $\overline{V}_o$ given by Eq. (\ref{4.19})) and $\dot{V}(0) = 0$.}
		\label{Fig. 9a}
	\end{subfigure}	
	
	\begin{subfigure}{.5\textwidth}
		\includegraphics[width=\linewidth]{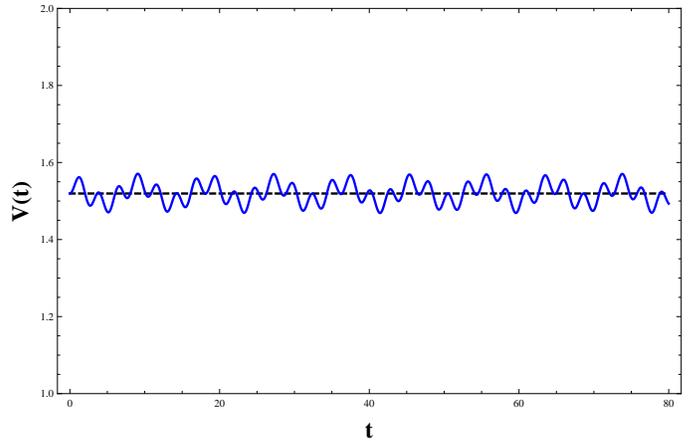}	
		\caption{Plot of $V(t)$ vs $t$ with $V(0) = V_o = 1.51949$ and $\dot{V}(t) = 0$.}
		\label{Fig. 9b}
	\end{subfigure}	
	\caption{Numerics of $\langle x \rangle$ and $V(t)$ with $\omega = 1$, $\lambda = 0.1$ and $e = -0.5$.}
	\label{Fig. 9}
	\end{figure}

	Now Eq. (\ref{4.17}) is always satisfied and so is Eq.(\ref{4.18}) and hence these fixed points are always stable. The numerics of $\langle x \rangle$ and $V$ around the fixed point $B$ is shown in Fig. \ref{Fig. 9}. 
	
	\vspace{.1in}	
	
	As a coupled dynamical system, this is fascinating since it shows that over a large energy range these are two stable fixed points and hence there will be two distinct basins of attraction and the basin boundary could be an interesting object of study. We forego that because the physical problem we started with should have oscillation from one well to another for all $E$ and the fact that there is a restriction on $E$ is a shortcoming of the Gaussian approximation. We end this section by pointing out that if we use a non-zero skewness, then Eq. (\ref{4.4}) and Eq. (\ref{4.5}) become
	
	\begin{subequations} \label{4.20}
	\begin{equation}
		\frac{d^2}{dt^2}\langle x \rangle = (\omega^2 - 3\lambda V)\langle x \rangle - \lambda \langle x \rangle^3 - \lambda S	
	\end{equation}		
	\begin{align}
		\frac{d^2}{dt^2}V = \ &4e + 4\omega^2 V - 9\lambda V^2 + 2\omega^2 \langle x \rangle^2 \nonumber \\
							&-12\lambda V \langle x \rangle^2 - \lambda \langle x \rangle^4 - 10\lambda S \langle x \rangle	\nonumber \\ &-2\langle p \rangle^2
	\end{align}
	\end{subequations}
	
	If we use the phenomenological form $S = \beta_1\langle x \rangle + \beta_2 \langle x \rangle^3$, then the above equations become

	\begin{subequations} \label{4.21}
	\begin{align}
		\frac{d^2}{dt^2}\langle x \rangle = (&\omega^2 - 3\lambda V - \lambda \beta_1)\langle x \rangle \nonumber \\ 
											 &- \lambda(1 + \beta_2)\langle x \rangle^3	
	\end{align}		
	\begin{align}
		\frac{d^2}{dt^2}V = \ &4e + 4\omega^2 V - 9\lambda V^2 + 2\omega^2 \langle x \rangle^2 \nonumber \\
							&-12\lambda V \langle x \rangle^2 - \lambda \langle x \rangle^4 \nonumber \\ &- 10\lambda(\beta_1 \langle x \rangle^2 + \beta_2 \langle x \rangle^4) -2\langle p \rangle^2
	\end{align}
	\end{subequations}	
	
	We focus on the fixed point about $\langle x \rangle = 0$ and $V = V_o$. We still have $V_o$ given by Eq. (\ref{4.10}). The stability equations for linearisation about $V_o$ is still given by Eq. (\ref{4.9}) and stable for the reason given in the discussion following it. For linearisation around $\langle x \rangle = 0$, we now have
	
	\begin{equation} \label{4.22}
		\frac{d^2}{dt^2}\delta x = (\omega^2 - 3\lambda V_o - \lambda \beta_1)\delta x
	\end{equation}
	
	and stability requires that 
	
	\begin{equation} \label{4.23}
		(3V_o + \beta_1)\lambda > \omega^2
	\end{equation}
	
	\vspace{.1in}	
	
	Using Eq. (\ref{4.10}) this leads to, with $\beta_1 = \beta_0\omega^2$, $1 - \lambda \beta_0 < \frac{2}{3}[1 + \sqrt{1 + \frac{9\lambda e}{\omega^4}}]$, which for $\beta_o = \frac{1}{3\lambda}$ allows the origin to be stable for $e > -\frac{\omega^4}{9\lambda}$, i.e. over the entire range of existence of the fixed point.
	
	\vspace{.1in}	
	
	In studying the tunnelling in the double well, we found that out treatment of quantum fluctuations is particularly good for energies $E > -\frac{\omega^4}{12\lambda}$. This enables us to study the very short effect of quantum fluctuations in a completely different context - the effect on classical chaotic dynamics. The double well potential dynamics becomes chaotic in classical mechanics if we add a small amount of sinusoidal oscillation upto linear terms in the potential i.e. the potential is now 
	
	\begin{equation} \label{4.24}
		V(x,t) = -\frac{1}{2}\omega^2 x^2 + \frac{\lambda}{4} x^4 + g x \cos{\Omega t}
	\end{equation}	 
	
	The equation of motion in classical mechanics is 
	
	\begin{equation} \label{4.25}
		\ddot{x} = \omega^2 x - \lambda x^3 - g \cos{\Omega t}
	\end{equation}				
	
	The oscillatory term, as has been very well studied, induces chaos near the separatrix for $g = 0$. For $g = 0$, we can write down the integral of motion as,
	
	\begin{equation} \label{4.26}
		\omega (t - t_0) = \int_a^x \frac{dy}{\sqrt{\frac{2E}{\omega^2} + y^2 - \frac{\lambda}{2\omega^2} y^4}}	
	\end{equation}
	
\begin{figure}[h]
		\includegraphics[width=\linewidth]{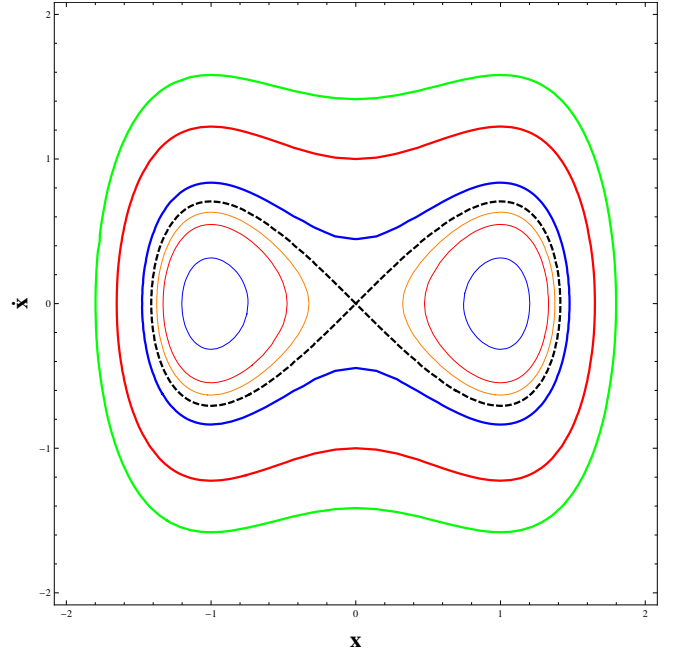}	
		\caption{(color online) Phase-space portrait for the double well potential with $\omega = 1$ and $\lambda = 1$ and different values of energy, $E$. The black dashed curve corresponds to $E = 0$, the separatrix for the double well potential. For $E < 0$, the particle oscillates about $\pm \sqrt{\frac{\omega^2}{\lambda}}$ depending on the initial conditions.}
		\label{Fig. 10}
	\end{figure}

	If the motion starts from $x = a$ with $\dot{x} = 0$ then $E = -\frac{1}{2}\omega^2 a^2 + \frac{1}{4} \lambda a^4$. For exactly $E = 0$, the trajectory can be easily found and has the form in the $x - \dot{x}$ plane as shown in Fig. \ref{Fig. 10}.

\begin{figure}[h]
	\begin{subfigure}{.5\textwidth}
		\includegraphics[width=\linewidth]{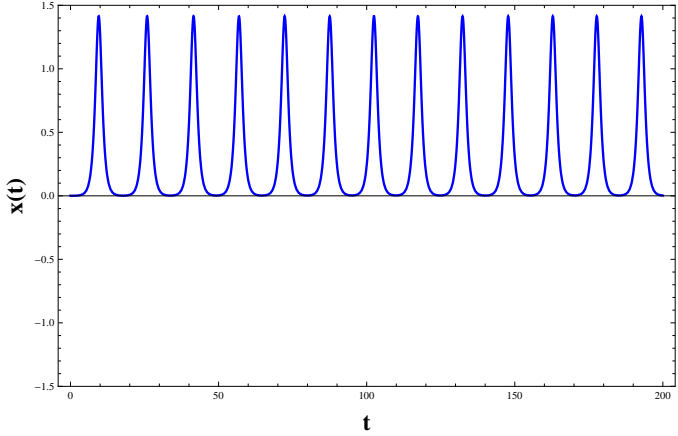}	
		\caption{Plot of $x(t)$ vs $t$ with $x(0) = .0014$ and $\dot{x}(0) = 0$.}
		\label{Fig. 11a}
	\end{subfigure}	
	
	\begin{subfigure}{.5\textwidth}
		\includegraphics[width=\linewidth]{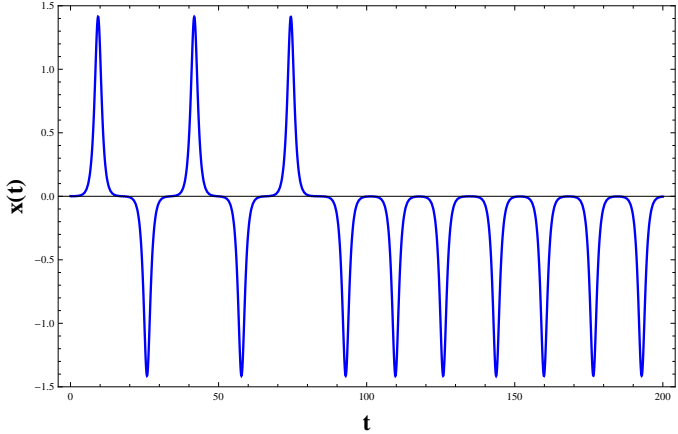}	
		\caption{Plot of $x(t)$ vs $t$ with $x(0) = .0015$ and $\dot{x}(0) = 0$.}
		\label{Fig. 11b}
	\end{subfigure}	
	\caption{Numerics for the potential Eq. (\ref{4.24}) with $\omega = 1$, $\lambda = 1$, $g = 0.1$ and $\Omega = 10$ for two closely spaced initial conditions and momentum $\dot{x} = 0$.}
	\label{Fig. 11}
	\end{figure}

	The trajectory $E = 0$(the homo-clinic) orbit separates two kinds of closed orbits, those with $E < 0$ and those with $E > 0$. If one switches on an oscillating potential at this stage, then a particle released with $E \simeq 0$ but negative(i.e. from within one of the wells) might see itself propelled into the other well by the oscillatory field but another released with a very slightly different initial conditions can find itself trapped in the same well leading automatically to a positive Lyapunov exponent. We show this scheme sensitivity to initial conditions in Fig. \ref{Fig. 11} where an initial condition of $x = .0014$, $\dot{x} = 0$ confines the dynamics in one of the wells while $x = .0015$ and $\dot{x} = 0$ confines the particle in the other well, thus ensuring a positive Lyapunov exponent.
	
	\begin{figure}
	\begin{subfigure}{.5\textwidth}
		\includegraphics[width=\linewidth]{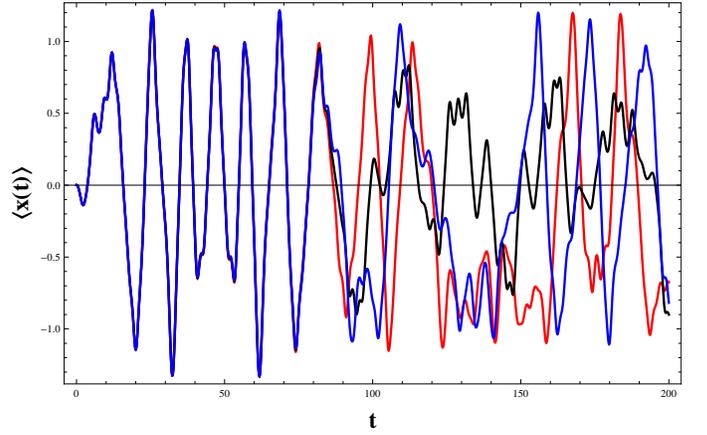}	
		\caption{Plot of $\langle x(t) \rangle$ vs $t$ with $V(0) = .10$}
		\label{Fig. 12a}
	\end{subfigure}	
	
	\begin{subfigure}{.5\textwidth}
		\includegraphics[width=\linewidth]{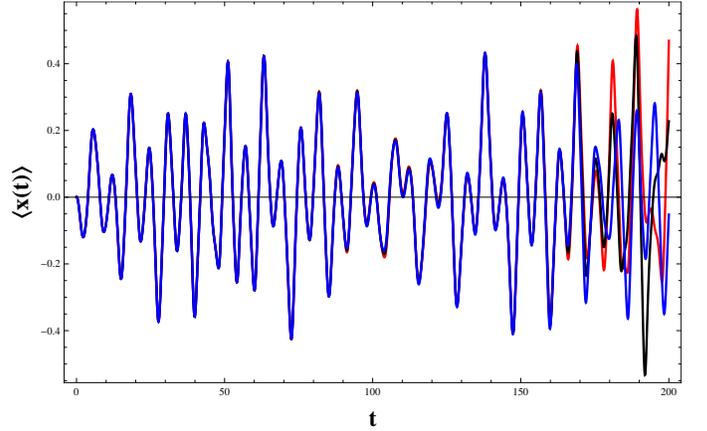}	
		\caption{Plot of $\langle x(t) \rangle$ vs $t$ with $V(0) = .25$}
		\label{Fig. 12b}
	\end{subfigure}
	\begin{subfigure}{.5\textwidth}
		\includegraphics[width=\linewidth]{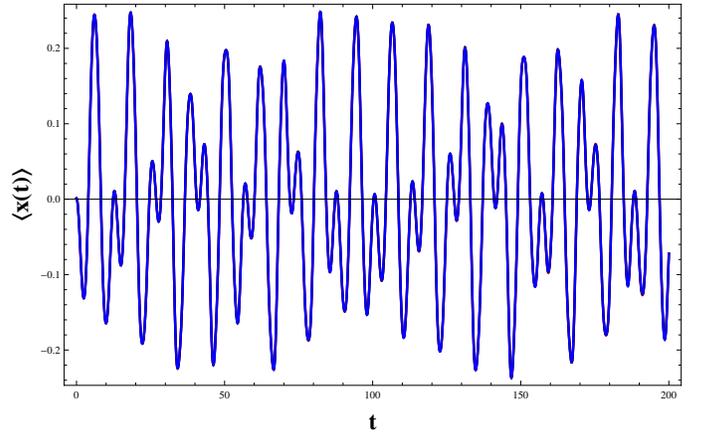}	
		\caption{Plot of $\langle x(t) \rangle$ vs $t$ with $V(0) = .40$}
		\label{Fig. 12c}
	\end{subfigure}	
	\caption{(color online) Numerics for the Eq. (\ref{4.27}) with $\omega = 1$, $\lambda = 1$, $g = 0.1$ and $\Omega = 1$ for three closely spaced initial conditions $\langle x(0) \rangle$ keeping $\langle \dot{x}(0) \rangle = 0$ and for different initial variance $V(0)$ keeping $\dot{V}(0) = 0$ starting at the separatrix $E(0) = 0$. The red, black and blue curve corresponds to $\langle x(0) \rangle = 0.0011, 0.0012$ and $0.0013$.}
	\label{Fig. 12}
	\end{figure}	
	
	We now consider the quantum dynamics where the system with the potential of Eq. (\ref{4.24}) is governed by the Schroedinger's equation $\imath \hbar \frac{\partial \psi}{\partial t} = [\frac{p^2}{2} + V(x,t)]\psi$. The corresponding dynamics for $\langle x \rangle$ and $V = \langle x^2 \rangle - \langle x \rangle^2$ (we work with $S = 0$) can be easily seen to be,(we note that $\langle H \rangle = \frac{\langle p^2 \rangle}{2} - \frac{1}{2}\omega^2 \langle x^2 \rangle + \frac{\lambda}{4}\langle x^4 \rangle + g \langle x \rangle \cos{\Omega t}$ is no longer a constant of motion and has to be considered as a dynamical variable)
	
	\begin{subequations} \label{4.27}
		\begin{equation}
			\frac{d^2 \langle x \rangle}{dt^2} = (\omega^2 - 3\lambda V)\langle x \rangle - \lambda \langle x \rangle^3 - g \cos{\Omega t}			
		\end{equation}	
		\begin{align}
			\frac{d^2 V}{dt^2} = \ &4\langle H \rangle + 4\omega^2 V - 9\lambda V^2 - 2\langle p \rangle^2 \nonumber \\ &+ 2\omega^2 \langle x \rangle^2 - 12\lambda V \langle x \rangle^2 - \lambda \langle x \rangle^4  \nonumber \\ &- 4g\langle x \rangle \cos{\Omega t} 	
		\end{align}
		\begin{equation}
			\frac{d \langle H \rangle}{dt} = -g \ \Omega \ \langle x \rangle \sin{\Omega t}	
		\end{equation}							
	\end{subequations}

	In the dynamics of $\langle x \rangle$, the existence of $V$ actually leads to a blurring of the separatrix and hence the main ingredient for chaos is lost. If we now look at the coupled dynamics of $\langle x \rangle$, $V$ and $\langle H \rangle$ given in Eq. (\ref{4.27}), the dynamics changes character as the initial value of $V$ is changed from zero(classical limit) to large values. This is apparent in Fig. \ref{Fig. 12}. The figure shows the numerics for $\langle x \rangle$ for three closely spaced initial conditions. As $V$ is increased from its zero(classical limit) to higher values, the dynamics becomes insensitive to the initial conditions for larger time. This is highlighted in Fig. \ref{Fig. 12}.
	
	The moment equations are thus capable of smoothing out the dynamics of $\langle x \rangle$ through the coupling to the variance and higher moments.
	
	\section{Conclusion}	
	
	In conclusion we have used the Heisenberg's equation of motion to write the quantum dynamics as a hierarchy of infinite set of coupled ordinary differential equations. We have used closure schemes to arrive at practical answers. This allows us to talk about approximately shape independent state for non-quadratic potentials, about escape from volcano like potentials and tunnelling in a double well potential. We could also see by considering the periodically driven double well that the separatrix chaos is smoothed out at long times by quantum fluctuations. Clearly the reduction to dynamical systems appears as an alternate way of studying quantum dynamics.

\end{document}